\documentclass[12pt]{article}
\usepackage{amsmath,amssymb,amsthm,amsxtra,overpic,bbm,bm,epsfig,ulem}
\usepackage{color}
\usepackage{cite}

\usepackage{hyperref}
\usepackage{url}
\usepackage{comment}

\usepackage{tikz-feynman}
\usepackage{tikz}
\usetikzlibrary{decorations.pathmorphing}
\usetikzlibrary{decorations.pathmorphing} 
\tikzset{graviton/.style={decorate, decoration={snake, amplitude=.4mm, segment length=1.5mm, pre length=.5mm, post length=.5mm}, double}}

\def\thefootnote{\fnsymbol{footnote}}

\begin{document}

\vspace{0.2cm}

\begin{center}
{\large\bf  Testing the type-II seesaw mechanism with gravitational waves}
\end{center}

\vspace{0.2cm}

\begin{center}
{\bf Yonghua Wang }$^{1,2}$
\footnote{E-mail: yonghuawang@mail.bnu.edu.cn}
{\bf Wei Chao }$^{1,2}$
\footnote{E-mail: chaowei@bnu.edu.cn}
\\
{\small $^{1}$Key Laboratory of Multi-scale Spin Physics, Ministry of Education, Beijing Normal University, Beijing 100875, China} \\
{\small $^{2}$Center of Advanced Quantum Studies, School of Physics and Astronomy, Beijing Normal University, Beijing, 100875, China} 
\end{center}

\vspace{1cm}

\begin{abstract}

Traditional seesaw mechanisms provide an elegant theoretical framework for explaining the small yet non-zero masses of neutrinos. Nevertheless, they face significant experimental challenges, primarily because the energy scale associated with the seesaw mechanism is too high to be directly probed in terrestrial experiments. In this paper, we explore the gravitational waves (GWs) generated via graviton bremsstrahlung during the decay of seesaw particles in the early Universe. Specifically, we compute the GW spectrum resulting from the decay of the Higgs triplet within the type-II seesaw model. Our results demonstrate that the resulting GW spectrum depends sensitively on the mass of the Higgs triplet and its couplings to the Standard Model Higgs doublet and the left-handed lepton doublet. The detection of such a high-frequency GW background could offer a unique experimental window into the seesaw mechanism and provide indirect evidence for its validity.


\end{abstract}

\newpage

\def\thefootnote{\arabic{footnote}}
\setcounter{footnote}{0}

\section{Introduction}

The origin of the tiny but non-zero neutrino masses is a longstanding puzzle in particle physics. The seesaw mechanism~\cite{Minkowski:1977sc,Yanagida:1979as,Gell-Mann:1979vob,Glashow:1979nm,Mohapatra:1979ia}, which extends the standard model (SM) with seesaw particles, such as right-handed neutrino singlets or scalar triplet with $Y=1$, has been so far the most elegant one to accommodate the active neutrino masses. Notably, the seesaw mechanism also provides explanation to the matter-antimatter asymmetry of the Universe via the Leptogenesis mechanism~\cite{Fukugita:1986hr}.  All those observations make the testing of the seesaw mechanism a topic deserving meticulous investigation. 

However, it is difficult to test the seesaw mechanism using the terrestrial experiments considering that the seesaw scale is approaching to the scale of the grand unification theory (GUT). Although many TeV scale seesaw mechanisms~\cite{Cai:2017jrq,Zee:1980ai,Babu:1988ki,Ma:2006km,Chao:2010fyp} have been proposed, they loss simplicity and naturalness in explaining the active neutrino masses. Given that a direct test is not possible, several indirect detection methods have been proposed, which includes examining  the lepton-flavor-violating process  in precision measurements~\cite{Lindner:2016bgg}, lepton number violation in neutrino-less double beta decay~\cite{Rodejohann:2011mu},  or testing Leptogenesis with cosmological collider~\cite{Cui:2021iie}. 

Since their discovery by ground-based interferometers~\cite{LIGOScientific:2016aoc}, gravitational waves (GWs) have become a new probe of the early-Universe physics, as they allow us to bypass the cosmic fog that obscures light and offer a direct window into epochs and physical processes—such as inflation~\cite{barman2023gravitational,Xu:2025wjq,huang2019stochastic,Barman:2023rpg,Bernal:2023wus,Xu:2024fjl,Xu:2024xmw,Bernal:2024jim,Bernal:2025lxp,Tokareva:2023mrt,Kanemura:2023pnv,Montefalcone:2025gxx,Ema:2020ggo,Bernal:2023wus,Klose:2022knn,Klose:2022rxh}, GUT~\cite{An:2022cce,Hu:2025xdt,Chao:2017ilw}, dark matter~\cite{Chao:2017vrq,Chao:2023lox,Wang:2025lmf,Konar:2025iuk}, and cosmic phase transitions~\cite{Caprini:2009yp,Chao:2017vrq,Hindmarsh:2017gnf,Athron:2023xlk,Chao:2021xqv}—that would otherwise be inaccessible.
There are proposals for detecting  GW spectrum in various frequency bands. Especially,  efforts are underway in designing new facilities for testing high frequency GW, which mainly arise from the physics at high energy scale.  It has been shown~\cite{Ghiglieri:2015nfa,Ringwald:2022xif,Ghiglieri:2022rfp,Ghiglieri:2024ghm,Ringwald:2020ist} that the thermal GW background, an analogue of the cosmic microwave background, 
reflects the energy distribution of the SM plasma form which they are produced by microscopic collisions.
Mapping these observations into neutrino physics,  one can conclude that GW can be taken as an indirect probe of the seesaw mechanism.  
Actually, people have attempted to observing Leptogenesis  with GWs spectrum form graviton bremsstrahlung in decays of right-handed neutrinos during Leptogenesis~\cite{datta2024probing,choi2025cosmological,murayama2025observing,Kanemura:2025rct}.  

In this paper, we study the GW spectrum induced from the type-II seesaw mechanism, which extends the SM with a $Y=1$ scalar triplet $\Delta$ that couples to left-handed lepton doublets $\ell_L $ as well as SM Higgs doublet $H$. Our results differ from the study of GWs  induced by the type-I seesaw model in the following ways: (1) there are two decay channels in the type-II seesaw model, namely $\Delta \to HH$ and $\Delta\to \ell_L \ell_L$, the interplay of which may explicitly modify the decoupling temperature of the $\Delta$ in the early Universe; (2) the GW spectrum from the type-II seesaw mechanism is relevant to the match between these two-type interactions. Our illustrative numerical results show that the GW spectrum induced by the bremsstrahlung in the decay of the Higgs triplet is accessible by the proposed cavity facilities~\cite{Herman:2020wao, Herman:2022fau}  which are designed for the direct detection of high frequency GWs.

The remaining the paper is organized as follows: In section II and III, we give a brief overview on the type-II seesaw mechanism and the decay of the Higgs triplet as well as its evolution in the early Universe. Section IV is devoted to the calculation of the GW spectrum. The last part is conclusion remarks. The detail of the calculation, including the Feynman rules and the squared amplitudes are given in the appendix A and B.

\section{The type-II seesaw model}

The type-II seesaw mechanism extends the SM with a  $SU(2)_L$ triplet scalar field $\Delta$ with hypercharge $Y=1$. The matrix representation of $\Delta$ reads as \cite{FileviezPerez:2008jbu}
\begin{equation}
    \Delta = \begin{pmatrix}
    \frac{\Delta^+}{\sqrt{2}} & \Delta^{++} \\
    \Delta^0 & -\frac{\Delta^+}{\sqrt{2}}
    \end{pmatrix}
\end{equation}
where $\Delta^{++}$, $\Delta^+$ and $\Delta^0$ are the doubly-charged, singly-charged and neutral components, whose charge conjugate states are denoted as $\Delta^{--}$, $\Delta^-$ and $\overline{\Delta^0}$, respectively. The relevant Lagrangian is given by
\begin{eqnarray}
{\cal L} = {\cal L}_{\rm SM}^{} + {\rm Tr} \left[  (D_\mu \Delta)^\dagger (D^\mu \Delta )\right] -V(H, \Delta)  - \frac{1}{2} Y_\Delta^{ij} \overline{\ell_{Li}^C} i \sigma_2 \Delta \ell_{Lj}^{} + {\rm h.c.} \label{master0}
\end{eqnarray}
where $D_\mu$ is the covariant derivative, $\ell_L$ is the left-handed lepton doublet with $Y_\Delta$ the $3\times 3$ symmetric Yukawa coupling between the Higgs triplet and lepton doublets, $C=i\gamma^2 \gamma^0$ being the charge conjugation operator.  $V(H, \Delta)$ is the potential taking the following form
\begin{eqnarray}
V(H, \Delta) &=& -\mu_H^2 H^\dagger H + \lambda (H^\dagger H)^4  + M_\Delta^2 {\rm Tr } [\Delta^\dagger \Delta^{} ] + \lambda_s {\rm Tr} [\Delta^\dagger \Delta]^2  + \lambda_{hs}^{} {\rm Tr } [\Delta^\dagger \Delta^{} ] (H^\dagger H ) \nonumber \\
&&- \frac{1}{2} \mu H^T i\sigma^2 \Delta^\dagger H + {\rm h.c.} \; , \label{master1}
\end{eqnarray}
where $M_\Delta^{}$ is the mass of the Higgs triplet, $\lambda$, $\lambda_s$ and $\lambda_{hs}$ are quartic couplings, $\mu$ is a dimension one parameter describing  triple coupling between the Higgs doublet and the triplet.  
After the electroweak symmetry breaking, the SM Higgs doublet acquires a vacuum expectation value (VEV) $v_0 = 246 \text{ GeV}$ \cite{ParticleDataGroup:2024cfk}.
The last term in the Lagrangian (\ref{master1}) induces  a VEV for the neutral component of the Higgs triplet, $v_\Delta^{}$. In the limit of $M_\Delta \gg v_0$, the VEV of the Higgs triplet can be approximated as
\begin{equation}
	v_\Delta \simeq \frac{\mu v_0 ^2}{4 M_\Delta ^2}.
\end{equation}
The neutrino mass matrix is then generated as
\begin{equation}
    M_\nu ^{ij} =Y_\Delta^{ij} v_{\Delta} = Y_\nu ^{ij} \frac{\mu v_0 ^2}{4 M_\Delta^2},
\end{equation}
which can be diagonalized by the the $3\times 3$ Pontecorvo-Maki-Nakagawa-Sakata (PMNS) matrix $\cal U$~\cite{Pontecorvo:1957cp,Maki:1962mu,Pontecorvo:1967fh},
\begin{equation}
	\mathcal{U}^\dagger M_\nu \mathcal{U}^* = \text{diag} \{m_1, m_2, m_3\}
\end{equation}
where $m_1$, $m_2$ and $m_3$ are the mass eigenvalues of active neutrinos, and $\cal U$ can be parameterized as
\begin{equation}
{\cal{U}} =
\begin{pmatrix}
c_{12} c_{13} &
s_{12} c_{13} &
s_{13} e^{-i\delta_{CP}} \\[6pt]
- s_{12} c_{23} - c_{12} s_{23} s_{13} e^{i\delta_{CP}} &
c_{12} c_{23} - s_{12} s_{23} s_{13} e^{i\delta_{CP}} &
s_{23} c_{13} \\[6pt]
s_{12} s_{23} - c_{12} c_{23} s_{13} e^{i\delta_{CP}} &
- c_{12} s_{23} - s_{12} c_{23} s_{13} e^{i\delta_{CP}} &
c_{23} c_{13}
\end{pmatrix}
\,
\mathrm{diag}\!\left(
e^{i\Phi_{1}/2},\,
1,\,
e^{i\Phi_{2}/2}
\right)
\label{eq:UPMNS}
\end{equation}
where $s_{ij} = \sin\theta_{ij}$, $c_{ij} = \cos\theta_{ij}$, $\delta_{CP}$ is the Dirac CP-violating phase, and $\Phi_1$ and $\Phi_2$ are the Majorana phases. 
Using data from from various neutrino oscillation experiments, the global fit results for neutrino masses and mixing parameters are presented in Table \ref{tab:neutrino_parameters} \cite{Li:2023ksw,Ansarifard:2022kvy,Esteban:2020cvm} for both normal hierarchy (NH) and inverted hierarchy (IH). 

\begin{table}[h]
    \centering
    \begin{tabular}{|c|c|c|}
        \hline
        Parameter & NH & IH \\ 
        \hline
        $\sin^2\theta_{12}$ & $0.304^{+0.012}_{-0.012}$ & $0.304^{+0.013}_{-0.012}$ \\ 
        $\theta_{12}/^\circ$ & $33.45^{+0.77}_{-0.75}$ & $33.45^{+0.78}_{-0.75}$ \\ 
        $\sin^2\theta_{23}$ & $0.450^{+0.019}_{-0.016}$ & $0.570^{+0.016}_{-0.022}$ \\ 
        $\theta_{23}/^\circ$ & $42.1^{+1.1}_{-0.9}$ & $49.0^{+0.9}_{-1.3}$ \\ 
        $\sin^2\theta_{13}$ & $0.02246^{+0.00062}_{-0.00062}$ & $0.02241^{+0.00074}_{-0.00062}$ \\ 
        $\theta_{13}/^\circ$ & $8.62^{+0.12}_{-0.12}$ & $8.61^{+0.14}_{-0.12}$ \\ 
		$\delta_{CP}/^\circ$ & $230^{+36}_{-25}$ & $278^{+22}_{-30}$ \\
        $\Delta m_{21}^2/10^{-5}~\mathrm{eV}^2$ & $7.42^{+0.21}_{-0.20}$ & $7.42^{+0.21}_{-0.20}$ \\ 
        $\Delta m_{3\ell}^2/10^{-3}~\mathrm{eV}^2$ & $+2.510^{+0.027}_{-0.027}$ & $-2.490^{+0.026}_{-0.028}$ \\ 
        \hline
    \end{tabular}
    \caption{Neutrino oscillation parameters with their best-fit values for both normal hierarchy (NH) and inverted hierarchy (IH) \cite{Ansarifard:2022kvy,Esteban:2020cvm}. $\Delta m_{3 \ell} = \Delta m_{31}$ for NH and $\Delta m_{3 \ell} = \Delta m_{32}$ for IH.}
    \label{tab:neutrino_parameters}
\end{table}
The recent DESI data \cite{DESI:2024mwx} prefers the normal hierarchy of neutrino masses \cite{Jiang:2024viw} and the lightest neutrino mass can be taken as massless, so the neutrino masses may be given as
\begin{equation}
	0 \sim m_1 \ll m_2 \approx \sqrt{\Delta m_{21}^2} \approx 8.6 \times 10^{-3} \text{ eV} < m_3 \approx \sqrt{\Delta m_{32}^2+\Delta m_{21} ^2} \approx 5.0 \times 10^{-2} \text{ eV} .
\end{equation}
Using the normal hierarchy data listed in Table \ref{tab:neutrino_parameters} and setting the Majorana phases to $\Phi_1=\Phi_2=0$ for simplicity, the Yukawa matrix $Y_\Delta$ is  given by
\begin{equation}
Y_\Delta \approx \frac{10^{-3} \text{eV}}{v_\Delta }
\left(
\begin{array}{ccc}
 2.4\, -1.1 i & -0.13+4.0 i & -6.0+4.4 i \\
 -0.13+4.0 i & 26\, +0.48 i & 21\, +0.074i \\
 -6.0+4.4 i & 21\, +0.074 i & 29\, -0.42i \\
\end{array}
\right).
\label{eq:Yukawa}
\end{equation}
There are constraints on the $v_\Delta$ from the precision measurements of the $\rho$-parameter \cite{ParticleDataGroup:2020ssz,Zhou:2022mlz} and ensuring that the observed neutrino masses are generated while maintaining the perturbative Yukawa couplings \cite{Barrie:2022cub}, which give the upper bound and lower bound.  One has 
\begin{equation}
	2.56 \text{ GeV} \gtrsim v_\Delta \gtrsim 0.05 eV .
\end{equation}
These bounds can be translated into constraints on the parameter $\mu$ as
\begin{equation}
	10 \text{GeV} \frac{M_\Delta ^2}{v_0 ^2} \gtrsim \mu \gtrsim 0.2 \text{eV} \frac{M_\Delta ^2}{v_0 ^2} .
\end{equation}
Given these constraints one can analyze the decay of the Higgs triplet, which will be done in the next section.

\section{Higgs triplet evolution in the early Universe}

\begin{figure}
	\centering
	\begin{minipage}
        [c]{.49\textwidth}
		\includegraphics[width=\textwidth]{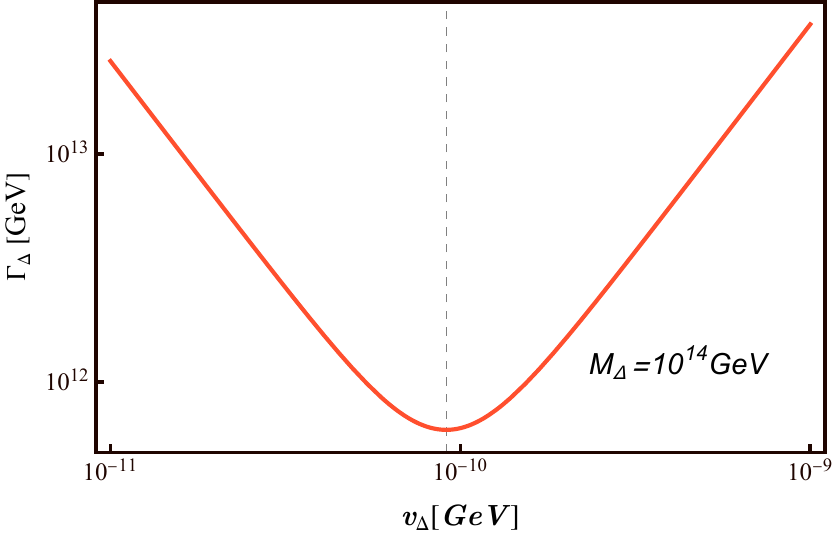}
	\end{minipage}
	\begin{minipage}
		[c]{.49\textwidth}
		\includegraphics[width=\textwidth]{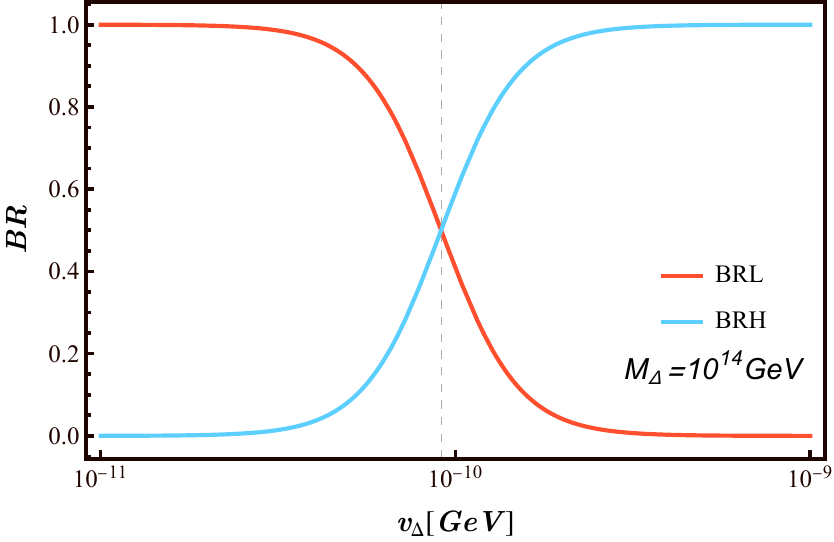}
	\end{minipage}
	\caption{
		\textbf{Left:} Decay widths of the Higgs triplet as functions of $v_\Delta$ for $M_\Delta = 10^{14}\,\text{GeV}$. \textbf{Right:} Branching ratios for Higgs-triplet decays into leptons (BRL) and into Higgs bosons (BRH) as functions of $v_\Delta$ for the same mass. The vertical dashed line in the left panel marks the value of $v_\Delta$ at which the total decay width is minimized; this value coincides with the intersection of BRL and BRH in the right panel. }
	\label{fig:BR}
\end{figure}

Given Lagrangian (\ref{master0}), we can directly extract the Feynman rules for various vertices, which are presented in the Appendix (\ref{Appendix:FeynmanRules}), and use them to calculate the decay widths of the Higgs-triplet components.
Decay widths for the di-lepton channel are given by\cite{Chun:2003ej,FileviezPerez:2008jbu,Li:2023ksw}
\begin{equation}
    \begin{aligned}
        \Gamma_{\Delta ^{++} \to e_i^{+} e_j^{+}} = \frac{\left| Y_\Delta ^{ij} \right|^2}{\delta_{ij}+1} \frac{M_\Delta}{16 \pi}, \quad
        \Gamma_{\Delta ^{+} \to e_i^{+} \bar{\nu}_j} = \left| Y_\Delta ^{ij} \right|^2 \frac{M_\Delta}{32 \pi}, \quad
        \Gamma_{\Delta ^{0} \to \bar{\nu}_i \bar{\nu}_j} = \frac{\left| Y_\Delta ^{ij} \right|^2}{\delta_{ij}+1} \frac{M_\Delta}{16 \pi}
    \end{aligned}
\end{equation}
and decay widths for the di-Higgs channel are given by
\begin{equation}
    \begin{aligned}
        \Gamma_{\Delta^{+ +} \to H^{+}H^{+}} = \frac{\mu^2}{32 \pi M_\Delta}, 
		\quad
        \Gamma_{\Delta^{+} \to H^{+}H^{0}} = \frac{\mu^2}{32 \pi M_\Delta}, 
		\quad
       \Gamma_{\Delta^{0} \to H^{0}H^{0}} = \frac{\mu^2}{32 \pi M_\Delta}
    \end{aligned}
\end{equation}
where $\delta_{ij}$ is the Kronecker delta function. The  charge conjugate partner of the Higgs triplet components shares the same decay width. The total decay width, summed  over rates for all decay channels, is given by
\begin{equation}
	\begin{aligned}
		\Gamma_{\Delta^{}} &= \sum_{i \ge j} \Gamma _{\Delta^{} \to \ell_{L_i}^{} \ell_{L_j}^{}} + \Gamma_{\Delta^{} \to H^{}H^{}} 
		= \sum_{i,j} \left| Y_\Delta ^{ij} \right|^2 \frac{M_\Delta}{32 \pi} + \frac{\mu^2}{32 \pi M_\Delta}\\
		&=\frac{M_\Delta}{32\pi} \left( \frac{2.57 \times 10^{-3} \text{eV}^2}{v_\Delta^2}  + \frac{16 M_\Delta^2 v_\Delta^2}{v_0^4}\right),
	\end{aligned}
\end{equation}
where $i$ and $j$ run from 1 to 3. In the first row, the summation over $i$ and $j$ is restricted to $i \ge j$ to avoid double counting for symmetric final states under the exchange of $i$ and $j$. In the second row, we have used  numerical results given in the Eq. (\ref{eq:Yukawa}). 
Branching ratios are defined as
\begin{equation}
	\text{BRL} = \frac{\sum_{i \ge j} \Gamma _{\Delta^{} \to \ell_{L_i}^{} \ell_{L_j}^{}}}{\Gamma_{\Delta}}, \quad \text{BRH} = \frac{\Gamma_{\Delta^{} \to H^{}H^{}} }{\Gamma_\Delta} .
\end{equation}
We plot the total decay width and the branching ratios as functions of $v_\Delta$ for fixed value of $M_\Delta$ in the left-panel of the Fig. \ref{fig:BR}. 
It shows that the branching ratios for the di-lepton and di-Higgs channels become equal and the total decay width is minimized when
\begin{equation}
	 v_\Delta =  \sqrt{\frac{0.01 \text{eV}}{M_\Delta}} v_0.
	 	 \label{mueq}
\end{equation}
We show in the right-panel of the Fig.~\ref{fig:BR} two branching ratios as the function of $v_\Delta$ by setting $M_\Delta^{} =10^{14}~{\rm GeV}$, which shows that two branching ratios equal with each other at $v_\Delta \sim 7.8\times 10^{-2}$~{\rm eV}.

Now we can evaluate the evolution of the Higgs triplets in the early Universe. 
Assuming that the mass of the Higgs triplet is smaller than the reheating temperature, $M_\Delta < T_{rh}$, the  Higgs triplet is initially relativistic and remain in thermal equilibrium with the SM particles through gauge interactions.
After reheating, the number density of the Higgs triplet is given by~\cite{Kolb:1990vq,Hooper:2024avz}
\begin{equation}
	n_\Delta^{} (T) = \frac{g_*^{\Delta} \zeta(3)}{\pi^2} T_{}^3,
\end{equation}
where $\zeta(3) \approx 1.202$ is the Riemann zeta function, and $g_* ^\Delta = 6$ being the relativistic effective degrees of freedom of the Higgs triplet.
%
%
%
Introducing variables $Y_\Delta \equiv n_\Delta / S$ and $x \equiv M_\Delta / T$, where $S = \frac{2\pi^2}{45} g_*^S(T) T^3$ being the entropy density and $g_*^S(T)$ is the relativistic effective degrees of freedom for entropy density at the temperature $T$, the Boltzmann equation for the Higgs triplets can be written as
\begin{equation}
	\frac{dY_\Delta}{dx} = - \frac{\langle\Gamma_\Delta\rangle}{H x} \left( Y_\Delta - Y_\Delta^{eq} \right).
	\label{eq:BEQY}
\end{equation}
The thermally averaged decay width in this expression is given by
\begin{equation}
	\langle \Gamma_\Delta \rangle = \Gamma_\Delta \frac{K_1(M_\Delta/T)}{K_2(M_\Delta/T)},
\end{equation}
where $K_1$ and $K_2$ are the modified Bessel functions of the second kind. The equilibrium number density of the Higgs triplets is denoted by $n_\Delta^{eq}$. The Hubble parameter during the radiation-dominated era is given by 
\begin{equation}
	H(T) = \sqrt{\frac{g_*^\rho(T) \pi^2}{90}} \frac{T^2}{M_P},
\end{equation}
where $M_P=2.4 \times 10^{18} \text{ GeV}$ being the reduced Planck mass, and $g_*^\rho(T)$ is the relativistic effective degrees of freedom for energy density at temperature $T$.


\begin{figure}[t]
	\centering
	\includegraphics[width=0.6\textwidth]{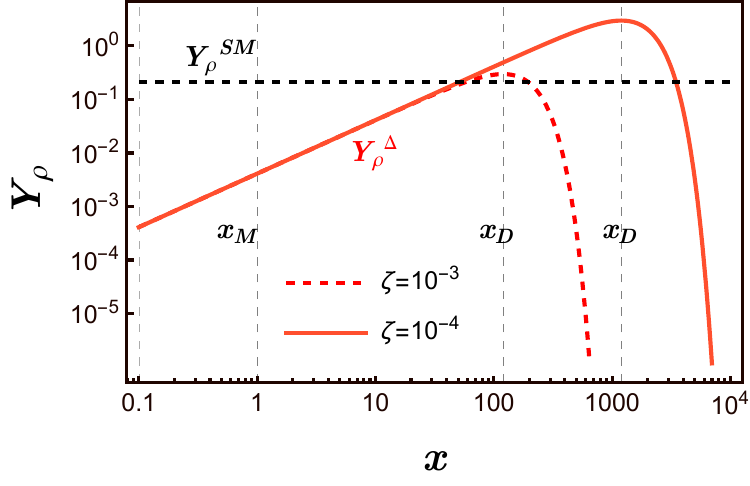}
	\caption{Evolution of the comoving energy densities for the Higgs triplet and radiation, with $M_\Delta = 10^{14} \text{ GeV}$ and $\zeta=10^{-4}$. The red solid line represents the comoving energy density of the Higgs triplets, $Y_\rho ^\Delta$, while the black horizential line represents that of radiation, $Y_\rho ^{SM}$.}
	\label{fig:Y_rho}
\end{figure}

The temperature at which the Higgs triplet decouples can be estimated by the following equation:
\begin{equation}
	H(T_D) = \langle \Gamma_\Delta \rangle .
	\label{eq:decaytime}
\end{equation}
If the neutrino masses are solely generated by the type-II seesaw mechanism, the total decay width is minimized when the branching ratios for the di-leptonic and di-Higgs channels equal with each other, resulting in the lowest decouple temperature. 
In this case, the parameters $v_\Delta$ and $M_\Delta$ are related with each other by Eq.(\ref{mueq}), leaving only a single free parameter, which can be chosen as $M_\Delta$.
Consequently, the Boltzmann equation in Eq.(\ref{eq:BEQY}) becomes independent of the model parameters. In this case, the decay temperature $T_D$ is always larger than the mass of the Higgs triplet for all values of $\mu$ and $M_\Delta$. 
Alternatively, if the neutrino masses are not solely generated by the type-II seesaw mechanism, the Yukawa couplings $Y_\Delta$ can not be soley determined by the Eq.(\ref{eq:Yukawa}). In this case,  we can redefine the total decay rate as 
\begin{equation}
	\Gamma_\Delta^{\rm tot} =\frac{ \zeta^2}{32 \pi} M_\Delta.
\end{equation}
where $\zeta$ is the function of the Yukawa coupling and the triple Higgs coupling constant. 
In this case, the Higgs triplets can decay after they become non-relativistic as long as 
$\zeta < 18 \sqrt{{M_\Delta}/{M_P}}$.
Furthermore, the universe may experience a early matter-dominated era due to the existence of non-relativistic Higgs triplets, if the energy density of the Higgs triplets $\rho_\Delta$ exceeds that of radiation $\rho_R$ before they decay. The condition for this situation to occur is 
\begin{equation}
	\zeta < 0.38 \sqrt{\frac{M_\Delta}{M_P}} \; .
\end{equation}
Figure~\ref{fig:Y_rho} shows the evolution of the comoving variable, $Y_\rho = \rho/S^{4/3}$, of the Higgs triplets and radiation for $M_\Delta = 10^{14}\,\text{GeV}$ with $\zeta = 10^{-3}$ and $10^{-4}$, where the contribution of the Higgs triplet decays to the radiation energy density has been neglected. 
As  can be seen from the figure, the Universe undergoes an early matter-dominated phase driven by the non-relativistic Higgs triplets before they decay out.

\section{Gravitational wave spectrum}

There are many sources for stochastic GWs in the early Universe, such as first order phase transitions\cite{Caprini:2015zlo,Caprini:2019egz,Hindmarsh:2020hop,Gouttenoire:2022gwi}, oscillation of cosmic strings\cite{Servant:2023tua}, or thermal fluctuations of the plasma, etc. In the early Universe, gravitons can also be generated from graviton bremsstrahlung process\cite{Weinberg:1965nx,Holstein:2006bh} in the  decays of the Higgs triplet. 
The Feynman diagrams that contribute to non-zero amplitudes of the graviton bremsstrahlung \cite{barman2023gravitational}  are shown in Fig. \ref{fig:Triplet_decay_diagram}. The detailed Feynman rules are presented in the Appendix (\ref{Appendix:FeynmanRules}).

\begin{figure}[t]
	\centering
\begin{minipage}[c]{.24\textwidth}
		\begin{tikzpicture}
			\begin{feynman}
				\vertex (a) {\(\Delta\)};
				\vertex [right=1.1cm of a] (b) ;
				\vertex [above right=1.1cm of b] (c);
				\vertex [above right=1.1cm of c] (d) {\(H\)};
				\vertex [below right=1.1cm of c] (e) {\(g \)};
				\vertex [below right=1.1cm of b] (f) {\(H \)};
				\diagram* {
					(a) -- [scalar,thick] (b), 
					(b) -- [scalar,thick] (c),
					(c) -- [scalar,thick] (d),
					(c) -- [graviton,thick] (e),
					(f) -- [scalar,thick] (b)
				};
				~~~~~~\end{feynman}
		\end{tikzpicture}
\end{minipage}	
	\begin{minipage}[c]{.24\textwidth}
		\begin{tikzpicture}
			\begin{feynman}
				\vertex (a) {\(\Delta\)};
				\vertex [right=1.1cm of a] (b) ;
				\vertex [above right=1.1cm of b] (c) {\(H\)};
				\vertex [below right=1.1cm of b] (d) ;
				\vertex [below right=1.1cm of d] (f) {\(H \)};
				\vertex [above right=1.1cm of d] (e) {\(g \)};
				\diagram* {
					(a) -- [scalar,thick] (b), 
					(b) -- [scalar,thick] (c),
					(d) -- [scalar,thick] (b),
					(d) -- [graviton,thick] (e),
					(f) -- [scalar,thick] (d)
				};
				~~~~~~\end{feynman}
		\end{tikzpicture}
	\end{minipage}
\begin{minipage}[c]{.24\textwidth}
		\begin{tikzpicture}
			\begin{feynman}
				\vertex (a) {\(\Delta\)};
				\vertex [right=1.1cm of a] (b) ;
				\vertex [above right=1.1cm of b] (c);
				\vertex [above right=1.1cm of c] (d) {\(\bar{L}\)};
				\vertex [below right=1.1cm of c] (e) {\(g \)};
				\vertex [below right=1.1cm of b] (f) {\(\bar{L}\)};
				\diagram* {
					(a) -- [scalar,thick] (b), 
					(c) -- [fermion,thick] (b),
					(d) -- [fermion,thick] (c),
					(c) -- [graviton,thick] (e),
					(f) -- [fermion,thick] (b)
				};
				~~~~~~\end{feynman}
		\end{tikzpicture}
\end{minipage}	
	\begin{minipage}[c]{.24\textwidth}
		\begin{tikzpicture}
			\begin{feynman}
				\vertex (a) {\(\Delta\)};
				\vertex [right=1.1cm of a] (b) ;
				\vertex [above right=1.1cm of b] (c) {\(\bar{L}\)};
				\vertex [below right=1.1cm of b] (d) ;
				\vertex [below right=1.1cm of d] (f) {\(\bar{L}\)};
				\vertex [above right=1.1cm of d] (e) {\(g \)};
				\diagram* {
					(a) -- [scalar,thick] (b), 
					(c) -- [fermion,thick] (b),
					(d) -- [fermion,thick] (b),
					(d) -- [graviton,thick] (e),
					(f) -- [fermion,thick] (d)
				};
				~~~~~~\end{feynman}
		\end{tikzpicture}
	\end{minipage}	
	\caption{Feynman diagrams for graviton bremsstrahlung in the decay of Higgs triplets. The left two diagrams are for the process $\Delta \to HHg$, and the right two diagrams are for the process $\Delta \to LLg$.}
	\label{fig:Triplet_decay_diagram}
\end{figure}
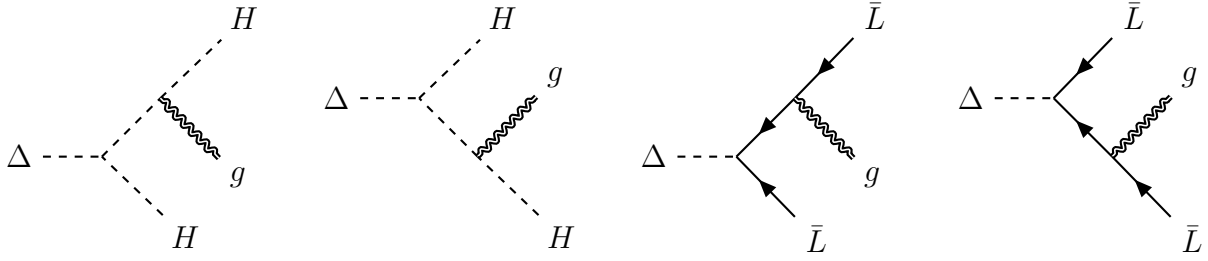

The evolution of the energy density of GWs is governed by the following Boltzmann equation
\begin{equation}
    \frac{d \rho_{GW}}{dt} + 4 H \rho_{GW} = \mathcal{C}_{\Delta \to LLg} + \mathcal{C}_{\Delta \to HHg}
\end{equation}
where $\mathcal{C}_{\Delta \to LLg}$ and $\mathcal{C}_{\Delta \to HHg}$ are the collision terms for generating graviton during the decay of the Higgs triplet into di-lepton and di-Higgs, given by
\begin{equation}
    \begin{aligned}
        \mathcal{C}_{\Delta \to LLg} &= \frac{3\kappa^2 M_\Delta \zeta^2}{512\pi^3}  \text{BRL}  \int d\omega n_\Delta \frac{K_1(M_\Delta/T)}{K_2(M_\Delta/T)} \left(1-2\frac{\omega}{M_\Delta}\right) \left(2-2\frac{M_\Delta}{\omega}+\frac{M_\Delta^2}{\omega^2}\right)\omega^2,
    \end{aligned}
\end{equation}
and
\begin{equation}
    \begin{aligned}
        \mathcal{C}_{\Delta \to HHg}
        &= \frac{3\kappa^2 M_\Delta \zeta^2}{512\pi^3} \text{BRH} \int d\omega  n_\Delta \frac{K_1(M_\Delta/T)}{K_2(M_\Delta/T)}  \left(2-\frac{M_\Delta}{\omega}\right)^2 \omega^2,
    \end{aligned}
\end{equation}
where $n_\Delta$ is the number density of the Higgs triplet, and $\kappa = \sqrt{2}/M_P$  being the coupling between graviton and SM particles.
The detailed derivation of the above two equations can be found in the Appendix (\ref{Appendix:CollisionTerms}).

Introducing the comoving variable $ Y_{GW} = \rho_{GW}/S^{4/3} $ and $x = M_\Delta / T$, the Boltzmann equation for the energy density of GWs can be written as
\begin{equation}
	\frac{d Y_{GW}}{dx} = \frac{1}{H x} S^{-4/3} \left( \mathcal{C}_{\Delta \to LLg} + \mathcal{C}_{\Delta \to HHg} \right). \label{masterx}
\end{equation} 
The GWs spectrum at present time is given by
\begin{equation}
    \Omega_{GW} h^2 (f) = \frac{1}{\rho_{cr,0}} \frac{d \rho_{GW,0}}{d \ln f} h^2  =\frac{1}{\rho_{cr,0}} \frac{d \left( Y_{GW}^{}(\infty) S(T_0)^{4/3} \right)}{d \ln f} h^2 
\end{equation}
where $\rho_{cr,0} = 3 H_0^2 M_P^2$ is the present-day critical energy density of the universe, with $H_0 = 100\, h\, \mathrm{km\, s}^{-1}\, \mathrm{Mpc}^{-1}$ being the Hubble parameter today. $\rho_{GW,0}$ denotes the present energy density of GWs, and $f$ is the present-day frequency of GWs, which is related to the graviton energy by $f = \omega_0 / 2\pi$, where $\omega_0$ is the graviton energy at present.

\begin{figure}[t]
	\centering
	\begin{minipage}
		[c]{.49\textwidth}
		\includegraphics[width=\textwidth]{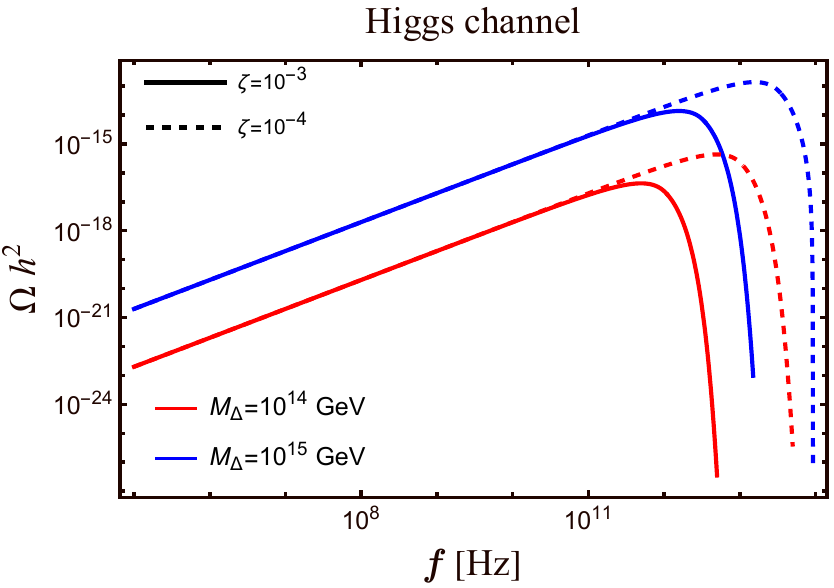}
	\end{minipage}
	\begin{minipage}
		[c]{.49\textwidth}
		\includegraphics[width=\textwidth]{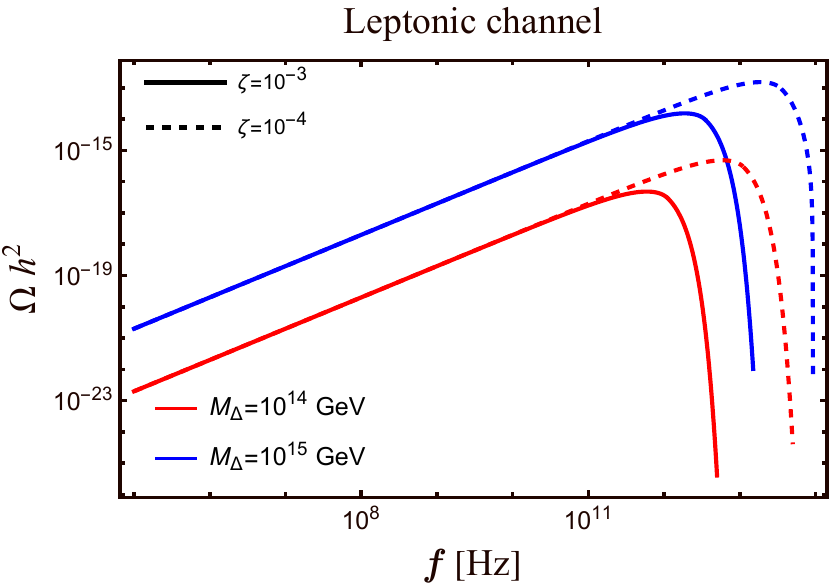}
	\end{minipage}
	\caption{\textbf{Left}: Gravitational wave spectrum produced by the Higgs decay channel for different values of $M_\Delta$ and $\zeta$. \textbf{Right}: Gravitational wave spectrum produced by the leptonic decay channel for different values of $M_\Delta$ and $\zeta$. The red and blue lines correspond to $M_\Delta = 10^{14} \text{ GeV}$ and $M_\Delta = 10^{15} \text{ GeV}$, respectively. The solid and dashed lines represent $\zeta = 10^{-3}$ and $\zeta = 10^{-4}$, respectively. }
	\label{fig:GW_spectrum}
\end{figure}

By solving the Boltzmann equation (\ref{masterx}), we  obtain the present-day GWs spectrum as
\begin{equation}
	\begin{aligned}
		\Omega_{GW} h^2 &= \frac{h^2}{\rho_{cr,0}} \frac{d \rho _{GW}}{d \ln \omega} (t_0) =\frac{h^2}{\rho_{cr,0}} S(T_0)^{4/3} \int_{x_{D}}^{x_0} dx  \frac{d}{d \ln\omega} Y_{GW} \\
		&= \frac{3 M_\Delta \zeta^2}{256\pi^3 M_p^2} \frac{h^2}{\rho_{cr,0}} S(T_0)^{4/3} \int_{x_{D}}^{x_0} dx \frac{M_\Delta^3}{H x} S^{-1/3}(x) Y_\Delta(x) \frac{K_1(x)}{K_2(x)} \frac{\omega(x)}{M_\Delta}  \\
		&\Big[\text{BRL} \left(1-2\frac{\omega(x)}{M_\Delta}\right) \left(1-2\frac{\omega(x)}{M_\Delta}+2\frac{\omega(x)^2}{M_\Delta^2}\right) +\text{BRH} \left(1-2\frac{\omega(x)}{M_\Delta}\right)^2 \Big],
	\end{aligned}
	\label{eq:spectrum}
\end{equation}
where the energy of the graviton is redshifted due to the expansion of the Universe as
\begin{equation}
	\omega(T) = \frac{a(T_0)}{a(T)} \omega_0=\left(
		\frac{g_*^S (T)}{g_*^S(T_0)}
	\right)^{1/3} \frac{T}{T_0} \omega_0.
\end{equation}

\begin{figure}[t]
	\centering
	\includegraphics[width=0.75\textwidth]{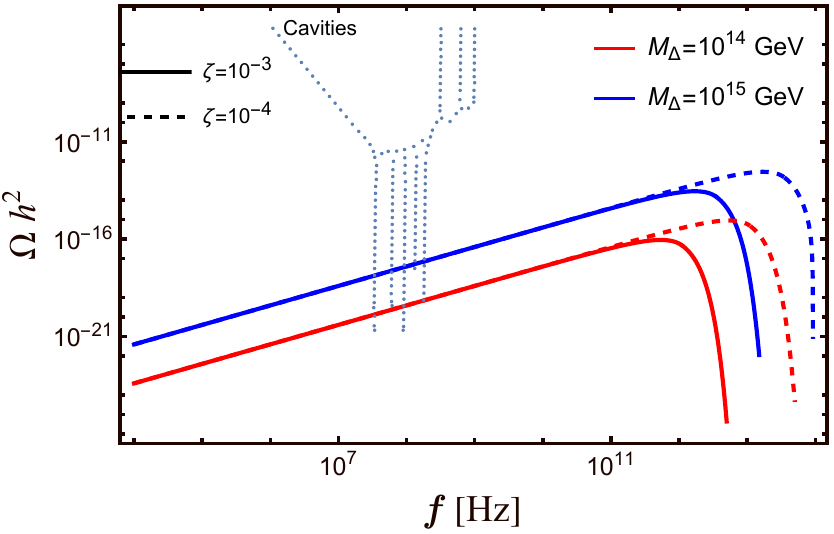}
	\caption{Gravitational wave spectrum produced by the decay of Higgs triplets with different values of $M_\Delta$ and $\zeta$, assuming equal branching ratios for the leptonic and Higgs channels, i.e., $\text{BRL} = \text{BRH} = 1/2$. The red and blue lines correspond to $M_\Delta = 10^{14} \text{ GeV}$ and $M_\Delta = 10^{15} \text{ GeV}$, respectively. The solid and dashed lines represent $\zeta = 10^{-3}$ and $\zeta = 10^{-4}$, respectively. The sensitivity curves for the resonant-cavity experiments are depicted with dotted lines.} 
	\label{fig:GW_spectrum_sensitivity}
\end{figure}

We show the GW spectrum arising from the bremsstrahlung in the decay of Higgs triplet in the Fig. \ref{fig:GW_spectrum}. The plot in the left panel shows the GW spectrum produced by the di-Higgs decay channel, while the plot in the right panel shows the GW spectrum produced by the di-lepton decay channel. The red and blue lines correspond to $M_\Delta = 10^{14} \text{ GeV}$ and $M_\Delta = 10^{15} \text{ GeV}$, respectively. The solid and dashed lines represent $\zeta = 10^{-3}$ and $\zeta = 10^{-4}$, respectively. 
In the Figure \ref{fig:GW_spectrum_sensitivity}, we show the total GW spectrum by setting branching ratios of the di-Higgs and di-lepton processes equal with each other.
As shown, the GW spectrum could be detected by the resonant cavities experiments~\cite{Herman:2020wao,Herman:2022fau,berlin2022detecting,berlin2023electromagnetic}.  Furthermore, future facilities for detecting higher frequency graviton may help to identify the signal of the Higgs triplet.

\section{Summary and remarks}

Seesaw are elegant mechanisms in addressing the tiny but nonzero active neutrino masses. Testing these mechanisms are longstanding challenges of the high energy physics.  People have proposed a new strategy to test the physics of the very Universe via high frequency GWs. In this paper, we have calculated the GW spectrum arising from the graviton bremsstrahlung process during the out of equilibrium decay of the Higgs triplet, which is the typical particle in the type-II seesaw model. We have studied the thermal history of the Higgs triplet, which may causes a pre-matter-dominated epoch in the early Universe for typical couplings between the Higgs triplet and the SM plasma. The GW spectrum is also derived, whose shape depends on the inputs of $Y_{\Delta}^{}$ and $\mu$. Although, currently there is no technology to observe GWs beyond kHz frequency, the GW spectrum emitted from the decay of the Higgs triplet can be reached by proposed cavity experiments designed for detecting high-frequency GWs.  Understanding the origin of high-frequency GWs is a new pathway to explore the high energy scale physics. Developing  new capability of measuring high frequency GWs may help us open the door of the neutrino physics  in the ultraviolet sector as well as the new physics beyond the SM.

\section*{Acknowledgements}

This work was supported by the National Natural Science Foundation of China (NSFC) (Grants No. 12447105, No. 11775025 and No. 12175027), and the Fundamental Research Funds for the Central Universities.

\appendix
\section{Feynman rules } \label{Appendix:FeynmanRules}
In this appendix, we present the Feynman rules for the relevant vertices used in our calculations.

\begin{figure}[htbp]
	\centering
	\begin{minipage}[c]{.32\textwidth}
		\begin{tikzpicture}
			\begin{feynman}
				\vertex (a) {\(\Delta^{++}\)};
				\vertex [right=2.2cm of a] (b) ;
				\vertex [above right=2.2cm of b] (c){\(H^+ \)};
				\vertex [below right=2.2cm of b] (d) {\(H^+ \)};
				\diagram* {
					(a) -- [scalar,thick, momentum=$P_1$] (b), 
					(b) -- [scalar,thick,momentum=$P_2$] (c),
					(b) -- [scalar,thick, momentum=$P_3$] (d)
				};
				~~~~~~\end{feynman}
		\end{tikzpicture}
	\end{minipage}	
	\begin{minipage}[c]{.32\textwidth}
		\begin{tikzpicture}
			\begin{feynman}
				\vertex (a) {\(\Delta^{+}\)};
				\vertex [right=2.2cm of a] (b) ;
				\vertex [above right=2.2cm of b] (c){\(H^+ \)};
				\vertex [below right=2.2cm of b] (d) {\(H^0\)};
				\diagram* {
					(a) -- [scalar,thick, momentum=$P_1$] (b), 
					(b) -- [scalar,thick,momentum=$P_2$] (c),
					(b) -- [scalar,thick, momentum=$P_3$] (d)
				};
				~~~~~~\end{feynman}
		\end{tikzpicture}
	\end{minipage}	
	\begin{minipage}[c]{.32\textwidth}
		\begin{tikzpicture}
			\begin{feynman}
				\vertex (a) {\(\Delta^{0}\)};
				\vertex [right=2.2cm of a] (b) ;
				\vertex [above right=2.2cm of b] (c){\(H^0 \)};
				\vertex [below right=2.2cm of b] (d) {\(H^0 \)};
				\diagram* {
					(a) -- [scalar,thick, momentum=$P_1$] (b), 
					(b) -- [scalar,thick,momentum=$P_2$] (c),
					(b) -- [scalar,thick, momentum=$P_3$] (d)
				};
				~~~~~~\end{feynman}
		\end{tikzpicture}
	\end{minipage}	
		\begin{minipage}[c]{.32\textwidth}
		\begin{tikzpicture}
			\begin{feynman}
				\vertex (a) {\(\Delta^{++}\)};
				\vertex [right=2.2cm of a] (b) ;
				\vertex [above right=2.2cm of b] (c){\(e_i^+ \)};
				\vertex [below right=2.2cm of b] (d) {\(e_j ^+\)};
				\diagram* {
					(a) -- [scalar,thick, momentum=$P_1$] (b), 
					(c) -- [fermion,thick] (b),
					(d) -- [fermion,thick] (b),
                    (b) -- [ momentum={$P_2$}] (c),
                    (b) -- [ momentum={$P_3$}] (d)
				};
				~~~~~~\end{feynman}
		\end{tikzpicture}
	\end{minipage}	
	\begin{minipage}[c]{.32\textwidth}
		\begin{tikzpicture}
			\begin{feynman}
				\vertex (a) {\(\Delta^{+}\)};
				\vertex [right=2.2cm of a] (b) ;
				\vertex [above right=2.2cm of b] (c){\(e_i^+\)};
				\vertex [below right=2.2cm of b] (d) {\(\bar{\nu}_j \)};
				\diagram* {
					(a) -- [scalar,thick, momentum=$P_1$] (b), 
					(c) -- [fermion,thick] (b),
					(d) -- [fermion,thick] (b),
                    (b) -- [ momentum={$P_2$}] (c),
                    (b) -- [ momentum={$P_3$}] (d)
				};
				~~~~~~\end{feynman}
		\end{tikzpicture}
	\end{minipage}	
    \begin{minipage}[c]{.32\textwidth}
		\begin{tikzpicture}
			\begin{feynman}
				\vertex (a) {\(\Delta^{0}\)};
				\vertex [right=2.2cm of a] (b) ;
				\vertex [above right=2.2cm of b] (c){\(\bar{\nu}_i \)};
				\vertex [below right=2.2cm of b] (d) {\(\bar{\nu}_j \)};
				\diagram* {
					(a) -- [scalar,thick, momentum=$P_1$] (b), 
					(c) -- [fermion,thick] (b),
					(d) -- [fermion,thick] (b),
                    (b) -- [ momentum={$P_2$}] (c),
                    (b) -- [ momentum={$P_3$}] (d)
				};
				~~~~~~\end{feynman}
		\end{tikzpicture}
	\end{minipage}	
	\caption{Feynman diagrams for the interaction vertices for the Higgs triplet and leptons/Higgs bosons. }
	\label{fig:Feynman_rules1}
\end{figure}
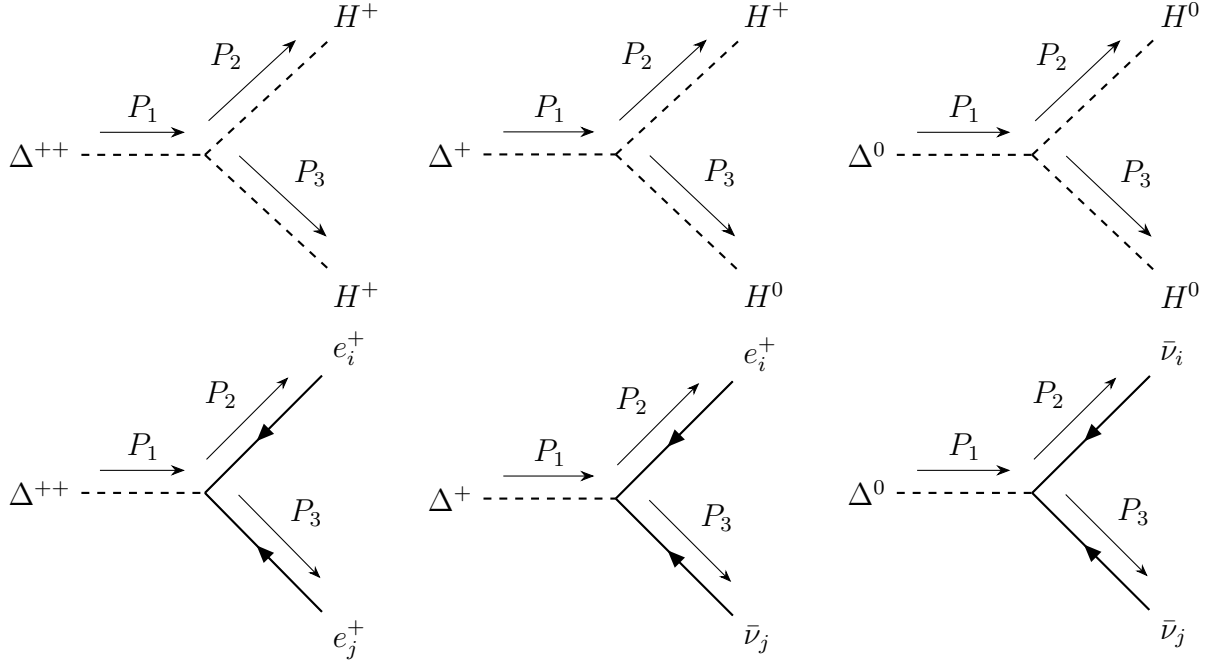

The interaction Lagrangian for the triplet and leptons/Higgs is given by Eq.(\ref{master0}), expanding it in terms of the component fields, we have \cite{Ma:1998dx}
\begin{equation}
\begin{aligned}
    \mathcal{L} \supset &-\frac{1}{2}\mu \left(\Delta^{--} H^+ H^+ - \sqrt{2} \Delta^{-} H^0 H^+ - \overline{\Delta^0} H^0 H^0\right)+h.c. \\
    & - \frac{1}{2}Y_\nu ^{ij} \left( \Delta^0 \nu_{iL}^T C \nu_{jL} - \sqrt{2} \Delta^+ e_{iL}^T C \nu_{jL} - \Delta^{++} e_{iL} ^T C e_{jL} \right) + h.c.
\end{aligned}
\end{equation}
where we have used the property that $Y_\nu^{ij} $ is symmetric.
The Feynman diagrams for the interaction vertices are shown in Fig.~\ref{fig:Feynman_rules1}, the corresponding vertex factors are:
\begin{equation}
    \begin{aligned}
        \Delta^{++} - H^+ - H^+ \text{vertex} &: - i \mu \\
        \Delta^{+} - H^0 - H^+ \text{vertex} &: i \frac{\sqrt{2}}{2} \mu \\
        \Delta^{0} - H^0 - H^0 \text{vertex} &:  i \mu \\
        \Delta^{++} - e_i ^+ - e_j^+ \text{vertex} &: -i  Y_\nu ^{ij} P_L \\
        \Delta^{+} - e_i^+ - \overline{\nu_j} \text{vertex} &: i \frac{\sqrt{2}}{2} Y_\nu ^{ij} P_L \\
        \Delta^{0} - \overline{\nu_i} - \overline{\nu_j} \text{vertex} &: -i  Y_\nu ^{ij} P_L 
    \end{aligned}
\end{equation}
where $P_L = (1-\gamma^5)/2$ is the left-handed projection operator. 

\begin{figure}[htbp]
    \centering
	\begin{minipage}{.24\textwidth}
        \begin{tikzpicture}
            \begin{feynman}
                \vertex (b) at (-2,-1) {\(\phi\)};
                \vertex (c) at (2,-1) {\(\phi\)};
                \vertex (d) at (0,2) {\(h_{\mu\nu}\)};
                \vertex [dot] (v) at (0,0) {};
                \diagram* {
                    (b) -- [scalar,thick,momentum=$P_1$] (v) -- [scalar,thick,momentum=$P_2$] (c),
                    (v) -- [graviton,thick] (d)
                };
            \end{feynman}
        \end{tikzpicture}
       \vspace{3pt}
    \footnotesize $- i\frac{\kappa}{2}\,[P_{1\mu}P_{2\nu}+P_{1\nu}P_{2\mu}]$
	\end{minipage}
	\hfill
	\begin{minipage}{.24\textwidth}
        \begin{tikzpicture}
            \begin{feynman}
                \vertex (b) at (-2,-1) {\(\chi\)};
                \vertex (c) at (2,-1) {\(\chi\)};
                \vertex (d) at (0,2) {\(h_{\mu\nu}\)};
                \vertex [dot] (v) at (0,0) {};
                \diagram* {
                    (b) -- [scalar,thick,momentum=$P_1$] (v) -- [scalar,thick,momentum=$P_2$] (c),
                    (v) -- [graviton,thick] (d)
                };
            \end{feynman}
        \end{tikzpicture}
	 \vspace{3pt}
    \footnotesize $- i\frac{\kappa}{2}\,[P_{1\mu}P_{2\nu}+P_{1\nu}P_{2\mu}]$
	\end{minipage}
	\hfill
	\begin{minipage}{.24\textwidth}
        \begin{tikzpicture}
            \begin{feynman}
               \vertex (b) at (-2,-1) {\(\psi\)};
                \vertex (c) at (2,-1) {\(\psi\)};
                \vertex (d) at (0,2) {\(h_{\mu\nu}\)};
                \vertex [dot] (v) at (0,0) {};
                \diagram* {
                    (b) -- [fermion,thick,momentum=$P_1$] (v) -- [fermion,thick,momentum=$P_2$] (c),
                    (v) -- [graviton,thick] (d)
                };
            \end{feynman}
        \end{tikzpicture}

	 \vspace{3pt}
    \footnotesize $-\frac{\kappa}{2}\frac{i}{4}\big[(P_1+P_2)_\mu\gamma_\nu+(P_1+P_2)_\nu\gamma_\mu\big]$
	\end{minipage}
	\caption{Feynman rules for the graviton interaction with real scalar, complex scalar and fermion fields. }
    \label{fig:graviton_vertex}
\end{figure}
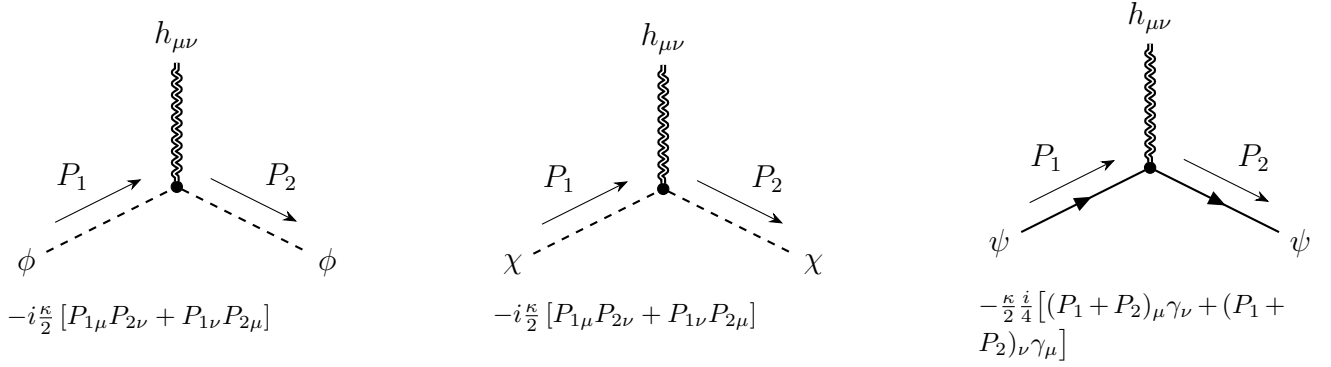

The graviton interacts with other particles through the following effective interaction Lagrangian:
\begin{equation}
	\mathcal{L}_{int} = -\frac{\kappa}{2} h_{\mu \nu} T^{\mu \nu}
\end{equation}
where \(h_{\mu \nu}\) is the graviton field, \(\kappa = \sqrt{2}/M_P \)~\cite{Nakayama:2018ptw}, and \(T^{\mu \nu}\) is the energy-momentum tensor of the matter fields. For the scalar field \(\phi\), comlex scalar field \(\chi\), and fermion field \(\psi\), their energy-momentum tensors are given by \cite{deAquino:2011ix}:
\begin{equation}
	\begin{aligned}
		T_{\mu \nu}^\phi &= \partial_\mu \phi \partial_\nu \phi - \eta_{\mu \nu} \mathcal{L}_\phi, \\
		T_{\mu \nu}^\chi &= \partial_\mu \chi^\dagger \partial_\nu \chi + \partial_\mu \chi \partial_\nu \chi^\dagger - \eta_{\mu \nu} \mathcal{L}_\chi, \\
		T_{\mu \nu}^\psi &= \frac{i}{4} \bar{\psi} \left( \gamma_\mu \overleftrightarrow{\partial}_\nu + \gamma_\nu \overleftrightarrow{\partial}_\mu \right) \psi - \eta_{\mu \nu} \mathcal{L}_\psi
	\end{aligned}
\end{equation}
where \( \mathcal{L}\) are the Lagrangians for the corresponding fields, and \(\bar{\psi}\overleftrightarrow{\partial}_\mu \psi \equiv \bar{\psi} \partial _\mu \psi - (\partial _\mu \bar{\psi }) \psi\). The polarization tensor of the graviton \(\varepsilon^{\mu\nu}\) satisfies the following conditions \cite{Gross:1968in,Gleisberg:2003ue}:
\begin{equation}
    \begin{aligned}
        \varepsilon ^{i \ \mu \nu} &= \varepsilon ^{i\ \nu \mu} \;\;\; \text{symmetric}\\
        \omega_\mu \varepsilon ^{i \ \mu \nu} &= 0 \;\;\;\; \; \;\;\;\text{transverse}\\
        \eta _{\mu \nu} \varepsilon ^{i \ \mu \nu} &= 0 \;\;\;\; \; \;\;\;\text{traceless}\\
        \varepsilon ^{i \ \mu \nu}\varepsilon ^{j\ *}_{\ \mu \nu} &= \delta ^{ij} \;\;\;\;\;\; \text{orthonormal}
    \end{aligned}
\end{equation}
where \(\omega_\mu\) is the four-momentum of the graviton. Due to the traceless condition, the non-zero interaction vertices are shown in Figure \ref{fig:graviton_vertex}. 

The polarization sum of the graviton is given by:
	\begin{equation}
    \sum_i \varepsilon _{i}^{\alpha \beta}\varepsilon _{i}^{* \mu \nu} = \frac{1}{2} \left( \hat{\eta}^{\alpha \mu}\hat{\eta}^{\beta \nu}+\hat{\eta}^{\alpha \nu}\hat{\eta}^{\beta \mu}-\hat{\eta}^{\mu \nu} \hat{\eta}^{\alpha \beta} \right)
\end{equation}
with 
\begin{equation}
    \hat{\eta}^{\mu \nu } \equiv \eta^{\mu \nu} - \frac{\omega^\mu \bar \omega ^\nu+\omega^\nu \bar \omega ^\mu}{\omega \cdot \bar \omega} 
\end{equation}
where \(\bar \omega ^\mu \equiv (\omega , - \vec{\omega})\). We use \( \omega \) to denote the energy of the graviton, \(\vec{\omega}\) to denote its spatial momentum, \(\omega^\mu\) to denote its four-momentum and \(\omega\cdot \bar{\omega}\) is the scalar product of two four-vectors.

\section{Collision terms} \label{Appendix:CollisionTerms}
In this section, we present the detailed derivation of the collision terms for generating GWs from the decay of Higgs triplets into a pair of leptons and a pair of Higgs bosons.

\subsection{Collision term for $\Delta \to LLg$}
The Higgs triplet has three components: a doubly charged scalar $\Delta^{++}$, a singly charged scalar $\Delta^{+}$, and a neutral scalar $\Delta^{0}$.
Each component can decay into a pair of leptons of any generation, accompanied by the emittion of a graviton. 
The Hermitian conjugate of the Higgs triplet undergoes similar graviton-emission processes involving antiparticles, which give the same contribution as their particle counterparts. 
The collision term for generating GWs from the leptonic decay channels of the Higgs triplet involves all these processes, and is given by the sum of the collision terms for each individual process:
\begin{equation}
    \mathcal{C}_{\Delta \to LLg} = 2 \left( \sum_{i\ge j}\mathcal{C}_{\Delta ^{++} \to e_i^+ e_j^+ g} + \sum_{ij}\mathcal{C}_{\Delta ^{+} \to e_i^+ \bar{\nu}_j g}+\sum_{i\ge j}\mathcal{C}_{\Delta ^{0} \to \bar{\nu}_i \bar{\nu}_j g}\right)
\end{equation}
where the factor of 2 comes from the contributions of the anti-particles. The summations over $i$ and $j$ run over the three lepton generations. For processes symmetric under $i\leftrightarrow j$, we sum only over $i\ge j$ to avoid double counting.

\begin{figure}[t]
	\centering
\begin{minipage}[c]{.40\textwidth}
		\begin{tikzpicture}
			\begin{feynman}
				\vertex (a) {\(\Delta^{++}\)};
				\vertex [right=1.1cm of a] (b) ;
				\vertex [above right=1.1cm of b] (c);
				\vertex [above right=1.1cm of c] (d) {\(e_i^+\)};
				\vertex [below right=1.1cm of c] (e) {\(g \)};
				\vertex [below right=1.1cm of b] (f) {\(e_j^+\)};
				\diagram* {
					(a) -- [scalar,thick] (b), 
					(c) -- [fermion,thick] (b),
					(d) -- [fermion,thick] (c),
					(c) -- [graviton,thick] (e),
					(f) -- [fermion,thick] (b)
				};
				~~~~~~\end{feynman}
		\end{tikzpicture}
\end{minipage}	
	\begin{minipage}[c]{.40\textwidth}
		\begin{tikzpicture}
			\begin{feynman}
				\vertex (a) {\(\Delta^{++}\)};
				\vertex [right=1.1cm of a] (b) ;
				\vertex [above right=1.1cm of b] (c) {\(e_i^+\)};
				\vertex [below right=1.1cm of b] (d) ;
				\vertex [below right=1.1cm of d] (f) {\(e_J^+ \)};
				\vertex [above right=1.1cm of d] (e) {\(g \)};
				\diagram* {
					(a) -- [scalar,thick] (b), 
					(c) -- [fermion,thick] (b),
					(d) -- [fermion,thick] (b),
					(d) -- [graviton,thick] (e),
					(f) -- [fermion,thick] (d)
				};
				~~~~~~\end{feynman}
		\end{tikzpicture}
	\end{minipage}	
	\caption{Feynman diagrams for graviton bremsstrahlung in the decay of double charged Higgs triplets.}
	\label{fig:double_charged_decay_diagram}
\end{figure}
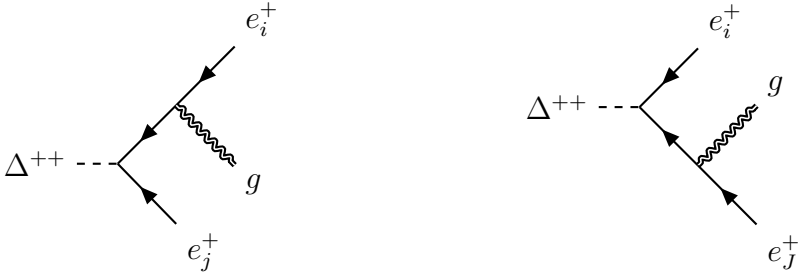

We present a detailed calculation of the collision term for $\Delta^{++}\to e_i^+ e_j^+ g$, and the collision terms for the other two processes can be obtained analogously. The collision term for $\Delta^{++}\to e_i^+ e_j^+ g$ is given by
\begin{equation}
    \begin{aligned}
        \mathcal{C}_{\Delta^{++} \to e_i^+ e_j^+ g} &=\int d\Pi_{\Delta^{++}} f_{\Delta^{++}} \frac{d\Pi_{e_i^+} d\Pi_{e_j^+} }{1+\delta_{ij}}d\Pi_g  (2\pi)^4 \delta^4 (\sum p_i - \sum p_f) | \mathcal{M}_{\Delta^{++} \to e_i^+ e_j^+ g} |^2 \omega.
    \end{aligned}
\end{equation}
Here, $ d \Pi= \frac{1}{(2\pi)^3 2E} d^3p $ is the Lorentz-invariant phase space volume element. The amplitude is the sum of the contributions from the two Feynman diagrams shown in Fig.~\ref{fig:double_charged_decay_diagram}, and is given by
\begin{equation}
    \begin{aligned}
         i \mathcal{M}_{\Delta^{++} \to e_i^+ e_j^+ g}  &= \bar{v}(p_{e_i^+}) \left(-i  Y_\Delta ^{ij} P_L\right) \frac{i\left(p_{e_j^+}+\omega\right)\cdot \gamma}{\left(p_{e_j^+}+\omega\right)^2} \left(-\frac{\kappa}{2} ip_{e_j^+}^\mu \gamma^\nu \right) v(p_{e_j^+}) \varepsilon_{\mu \nu}^* \\
        &+\bar{v}(p_{e_i^+})\left(-\frac{\kappa}{2} ip_{e_i^+}^\mu \gamma^\nu \right)\frac{i\left(p_{e_j^+}+\omega\right)\cdot \gamma}{\left(p_{e_j^+}+\omega\right)^2} \left(-i  Y_\Delta ^{ij} P_L\right)   v(p_{e_j^+}) \varepsilon_{\mu \nu}^*.
    \end{aligned}
\end{equation}
Here, the indices $i$ and $j$ are not summed. The Feynman diagram for emitting a graviton from the $\Delta^{++}$ line gives a vanishing contribution for the amplitude, because only moving particles can emit gravitons, and we choose the rest frame of $\Delta^{++}$ to calculate the Lorentz-invariant amplitude.

Squaring the amplitude and summing over the polarizations of the graviton and spins of the leptons, we have
\begin{equation}
    \begin{aligned}
        |\mathcal{M}_{\Delta^{++} \to e_i^+ e_j^+ g}|^2 &= \frac{\kappa^2}{8} |Y_\Delta ^{ij}|^2 M_\Delta ^2 \left(1-2\frac{\omega}{M_\Delta}\right) \left(2-2\frac{M_\Delta}{\omega}+\frac{M_\Delta^2}{\omega^2}\right).
    \end{aligned}
\end{equation}
The calculation of the amplitude was performed with the help of the package \texttt{FeynCalc}~\cite{mertig1991feyn,shtabovenko2016new,shtabovenko2020feyncalc,shtabovenko2025feyncalc}.

The integration over the phase space of final states gives
\begin{equation}
    \int d\Pi_{e_i^+} d\Pi_{e_j^+} d\Pi_g (2\pi)^4 \delta^4 (p_{\Delta^{++}} - p_{e_i^+} - p_{e_j^+} - p_g) = \int d\omega d E_{e_i^+} \frac{1}{32 \pi^3} 
\end{equation}

The distribution function of the Higgs triplet in thermal equilibrium is approximated by the Maxwell-Boltzmann distribution, i.e.,
$f_{\Delta^{++}} = e^{-E_{\Delta^{++}}/T}$. The phase space integration over $\Delta^{++}$ can be performed as
\begin{equation}
		\int d\Pi_{\Delta^{++}} f_{\Delta^{++}} = \frac{1}{2M_\Delta} n_{\Delta} \frac{K_1(M_\Delta/T)}{K_2(M_\Delta/T)},
\end{equation} 
where $n_{\Delta}$ is the number density of the Higgs triplet, $K_1$ and $K_2$ are the modified Bessel functions of the second kind, and we have assumed that the Higgs triplet remains in kinetic equilibrium while it decays.
With these ingredients, the collision term for $\Delta^{++} \to e_i^+ e_j^+ g$ can be written as
\begin{equation}
    \begin{aligned}
        \mathcal{C}_{\Delta^{++} \to e_i^+ e_j^+ g} &=\frac{\kappa^2 M_\Delta}{512\pi^3}\frac{|Y_\Delta ^{ij}|^2}{1+\delta_{ij}} \int d\omega d E_{e_i^+}n_{\Delta} \frac{K_1(M_\Delta/T)}{K_2(M_\Delta/T)} \left(1-2\frac{\omega}{M_\Delta}\right) \left(2-2\frac{M_\Delta}{\omega}+\frac{M_\Delta^2}{\omega^2}\right)\omega\\
        &= \frac{\kappa^2 M_\Delta}{512\pi^3}\frac{|Y_\Delta ^{ij}|^2}{1+\delta_{ij}} \int d\omega n_{\Delta} \frac{K_1(M_\Delta/T)}{K_2(M_\Delta/T)} \left(1-2\frac{\omega}{M_\Delta}\right) \left(2-2\frac{M_\Delta}{\omega}+\frac{M_\Delta^2}{\omega^2}\right)\omega^2.
    \end{aligned}
\end{equation}
In the second step, we integrated over $E_{e_i^+}$ from $M_\Delta /2 - \omega$ to $M_\Delta /2$. 

Similarly, we can obtain the collision terms for the other two processes:
\begin{equation}
    \begin{aligned}
        \mathcal{C}_{\Delta^{+} \to e_i^+ \bar{\nu}_j g} = \frac{\kappa^2 M_\Delta}{1024\pi^3}|Y_\Delta ^{ij}|^2\int d\omega  n_{\Delta} \frac{K_1(M_\Delta/T)}{K_2(M_\Delta/T)} \frac{\kappa^2}{16} \left(1-2\frac{\omega}{M_\Delta}\right) \left(2-2\frac{M_\Delta}{\omega}+\frac{M_\Delta^2}{\omega^2}\right)\omega^2,
    \end{aligned}
\end{equation}
and
\begin{equation}
    \begin{aligned}
        \mathcal{C}_{\Delta^{0} \to \bar{\nu}_i \bar{\nu}_j g} = \frac{\kappa^2 M_\Delta}{512\pi^3}\frac{|Y_\Delta ^{ij}|^2}{1+\delta_{ij}} \int d\omega n_{\Delta} \frac{K_1(M_\Delta/T)}{K_2(M_\Delta/T)} \left(1-2\frac{\omega}{M_\Delta}\right) \left(2-2\frac{M_\Delta}{\omega}+\frac{M_\Delta^2}{\omega^2}\right)\omega^2.
    \end{aligned}
\end{equation}
Summing over all the generations of leptons, we obtain the total collision term for $\Delta \to LLg$
\begin{equation}
	\mathcal{C}_{\Delta \to LLg} = \frac{3M_\Delta \kappa^2 }{512 \pi^3}\sum_{i,j} \left|Y_\Delta ^{ij}\right|^2 n_{\Delta}  \frac{K_1(M_\Delta/T)}{K_2(M_\Delta/T)} \int d\omega \left(1-2\frac{\omega}{M_\Delta}\right) \left(2-2\frac{M_\Delta}{\omega}+\frac{M_\Delta^2}{\omega^2}\right)\omega^2.
\end{equation}

\subsection{Collision term for $\Delta \to HHg$}
\begin{figure}[t]
	\centering
\begin{minipage}[c]{.24\textwidth}
		\begin{tikzpicture}
			\begin{feynman}
				\vertex (a) {\(\Delta^{++}\)};
				\vertex [right=1.1cm of a] (b) ;
				\vertex [above right=1.1cm of b] (c);
				\vertex [above right=1.1cm of c] (d) {\(H^+\)};
				\vertex [below right=1.1cm of c] (e) {\(g \)};
				\vertex [below right=1.1cm of b] (f) {\(H^+\)};
				\diagram* {
					(a) -- [scalar,thick] (b), 
					(b) -- [scalar,thick] (c),
					(c) -- [scalar,thick] (d),
					(c) -- [graviton,thick] (e),
					(f) -- [scalar,thick] (b)
				};
				~~~~~~\end{feynman}
		\end{tikzpicture}
\end{minipage}	
\hspace{1cm}
	\begin{minipage}[c]{.24\textwidth}
		\begin{tikzpicture}
			\begin{feynman}
				\vertex (a) {\(\Delta^{++}\)};
				\vertex [right=1.1cm of a] (b) ;
				\vertex [above right=1.1cm of b] (c) {\(H^+\)};
				\vertex [below right=1.1cm of b] (d) ;
				\vertex [below right=1.1cm of d] (f) {\(H^+\)};
				\vertex [above right=1.1cm of d] (e) {\(g \)};
				\diagram* {
					(a) -- [scalar,thick] (b), 
					(b) -- [scalar,thick] (c),
					(d) -- [scalar,thick] (b),
					(d) -- [graviton,thick] (e),
					(f) -- [scalar,thick] (d)
				};
				~~~~~~\end{feynman}
		\end{tikzpicture}
	\end{minipage}
	\caption{Feynman diagrams for graviton bremsstrahlung in the decay of doubly charged Higgs triplets into a pair of singly charged Higgs bosons.}
	\label{fig:double_charged_higgs_decay_diagram2}
\end{figure}
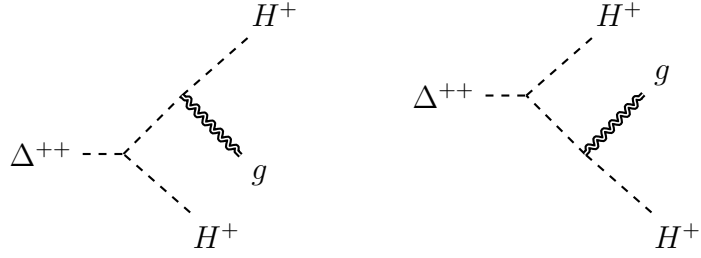

Simalarly, the collision term for generating GWs from the Higgs decay channels is given by 
\begin{equation}
    \begin{aligned}
        \mathcal{C}_{\Delta \to HHg} &=2\left( \mathcal{C}_{\Delta ^{++} \to H^+ H^+ g} + \mathcal{C}_{\Delta ^{+} \to H^0 H^+ g} + \mathcal{C}_{\Delta ^{0} \to H^0 H^0 g} \right)
    \end{aligned}
\end{equation}
where the factor of 2 comes from the contributions of the anti-particles.
In fact, the collision terms for these three processes are the same, we present a detailed calculation of the collision term for $\Delta^{++}\to H^+ H^+ g$, which is given by
\begin{equation}
    \begin{aligned}
        \mathcal{C}_{\Delta^{++} \to H^+ H^+ g} &=\frac{1}{2} \int d\Pi_{\Delta^{++}} f_{\Delta^{++}} d\Pi_{H^+_1} d\Pi_{H^+_2} d\Pi_g (2\pi)^4 \delta^4 (\sum p_i - \sum p_f) | \mathcal{M}_{\Delta^{++} \to H^+ H^+ g} |^2 \omega
    \end{aligned}
\end{equation}
Here, the subscripts 1 and 2 are used to distinguish the phase space variables of the two identical Higgs bosons in the final state, and the factor $1/2$ arises from the identical particles in the final states. 
The amplitude is the sum of the contributions from the two Feynman diagrams shown in Fig.~\ref{fig:double_charged_higgs_decay_diagram2}, and is given by
\begin{equation}
    \begin{aligned}
            i \mathcal{M}_{\Delta^{++} \to H^+ H^+ g}  &= \left(- i \mu\right) \frac{i}{(p_{H^+_2}+\omega)^2} \left(-\frac{\kappa}{2} i 2 p_{H^+_2}^\mu p_{H^+_2}^\nu \right) \varepsilon_{\mu \nu}^* \\
            &+\left(- i \mu\right) \frac{i}{(p_{H^+_1}+\omega)^2} \left(-\frac{\kappa}{2} i 2 p_{H^+_1}^\mu p_{H^+_1}^\nu \right) \varepsilon_{\mu \nu}^* 
    \end{aligned}
\end{equation}

Squaring the amplitude and summing over the polarizations of the graviton, we have
\begin{equation}
    \begin{aligned}
        |\mathcal{M}_{\Delta^{++} \to H^+ H^+ g}|^2 &= \frac{\kappa^2}{8} \mu^2 \left(2-\frac{M_\Delta}{\omega}\right)^2 
    \end{aligned}
\end{equation}
Therefore, the collision term for $\Delta^{++} \to H^+ H^+ g$ can be written as
\begin{equation}
    \mathcal{C}_{\Delta^{++} \to H^+ H^+ g} = \frac{1}{2} \frac{\kappa^2}{8} \mu^2 \frac{1}{32 \pi^3} \frac{n_{\Delta}}{ 2 M_\Delta} \frac{K_1(M_\Delta/T)}{K_2(M_\Delta/T)} \int d\omega  \left(2-\frac{M_\Delta}{\omega}\right)^2 \omega^2
\end{equation}
There is no a identical particles factor 1/2 in the collision terms for $ \Delta^{+} \to H^0 H^+ g$, but it's squared amplitude is half of that for $\Delta^{++} \to H^+ H^+ g$, so the collision term for $ \Delta^{+} \to H^0 H^+ g$ is the same as that for $\Delta^{++} \to H^+ H^+ g$, and same as the collision term for $\Delta^{0} \to H^0 H^0 g$. Thus, the total collision term for $\Delta \to HHg$ is given by
\begin{equation}
    \begin{aligned}
        \mathcal{C}_{\Delta \to HHg} &= 6 \mathcal{C}_{\Delta^{++} \to H^+ H^+ g} \\
        &= \frac{3}{512\pi^3}\kappa^2 \mu^2\frac{n_{\Delta}}{ M_\Delta}  \frac{K_1(M_\Delta/T)}{K_2(M_\Delta/T)}\int d\omega  \left(2-\frac{M_\Delta}{\omega}\right)^2 \omega^2
    \end{aligned}
\end{equation}

\bibliographystyle{elsarticle-num}
\bibliography{references}

\end{document}